\documentclass{article}

\usepackage{amssymb,amsfonts,amsmath}
\usepackage{cite,enumerate,float,indentfirst}
\usepackage{color}

\def\be{\begin{eqnarray}}
\def\ee{\end{eqnarray}}
\def\nn{\nonumber}

\def\Sch{{\rm Schur}}

\definecolor{red}{rgb}{1,0,0}
\definecolor{orange}{rgb}{1,0.5,0}
\definecolor{violet}{rgb}{0.7,0,1}

\hoffset= - 1.0in         

\begin{document}

\title{\vspace{-.5cm}{\Large {\bf  On Hopf-induced deformation of topological locus}\vspace{.2cm}}
\author{
{\bf A.Mironov$^{a,b,c}$}\footnote{mironov@lpi.ru; mironov@itep.ru}\ \ and
\ {\bf A.Morozov$^{b,c}$}\thanks{morozov@itep.ru}}
\date{ }
}

\maketitle

\vspace{-5cm}

\begin{center}
\hfill FIAN/TD-06/18\\
\hfill IITP/TH-08/18\\
\hfill ITEP/TH-10/18
\end{center}

\vspace{2.3cm}

\begin{center}
$^a$ {\small {\it Lebedev Physics Institute, Moscow 119991, Russia}}\\
$^b$ {\small {\it ITEP, Moscow 117218, Russia}}\\
$^c$ {\small {\it Institute for Information Transmission Problems, Moscow 127994, Russia}}

\end{center}

\vspace{.5cm}

\begin{abstract}
We provide a very brief review of the description of colored invariants for the Hopf link
in terms of characters, which need to be taken at
a peculiar deformation of the topological locus,
depending on one of the two representations associated with the
two components of the link.
Most important, we extend the description of this locus to
conjugate and, generically, to {\it composite} representations
and also define the ``adjoint" Schur functions
emerging in the dual description.
\end{abstract}

\vspace{.0cm}

\subsection*{Introduction}

The Hopf link

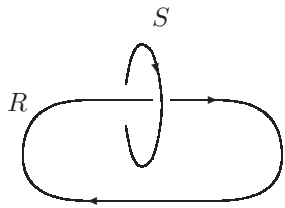
\begin{picture}(400,80)(-200,-50)

\put(0,2){\line(1,0){25}}
\put(32,2){\vector(1,0){18}}

\qbezier(25,-20)(32,0)(25,20)
\qbezier(25,-20)(18,-30)(15,-8)
\qbezier(25,20)(18,30)(15,8)



\qbezier(0,2)(-24,2)(-24,-19)
\qbezier(0,-36)(-24,-36)(-24,-19)
\put(50,-36){\vector(-1,0){50}}
\qbezier(50,2)(74,2)(74,-19)
\qbezier(50,-36)(74,-36)(74,-19)

\put(-30,-2){\mbox{${R}$}}
\put(25,30){\mbox{${S}$}}
\put(25,20){\vector(1,-4){2}}

\end{picture}

\noindent
is  the simplest non-trivial object in knot theory.
It plays a big role in various
branches of modern theory,
from the theory of the topological vertex \cite{topver}
to quantum computing \cite{quco}.


\bigskip

With Hopf link one associates a HOMFLY invariant ${\cal H}^{\rm Hopf}_{R\times S}= {\cal H}^{\rm Hopf}_{S\times R}\ $ \cite{HOMFLYPT},
which is a function of two variables $q$ and $A=q^N$ and two Young diagrams
${R}$ and ${S}$, which label finite-dimensional representations of the $sl_N$ algebra
associated with the two components of the link.
${\cal H}_{{R}\times{S}}^{\rm Hopf}(A,q)$ is actually a rational function,
but its denominator is a simple representation-dependent product of $q$-dependent
factors, and essential part is a Laurent polynomial in $A$ and $q$, which stands
in denominator. Hence, we often refer to the Hopf invariant as Hopf polynomial.
An exhaustive description of this polynomial is the first challenge in knot theory,
and this problem also provides the simplest relation to other branches of mathematical
physics, like the theory of topological vertices and topological models
associated with the resolved conifold.
We review the known solution to this problem for arbitrary representations given by $N$-independent Young diagrams
${S}$ and ${R}$ and extend it to the generic case when one of the representations
or both are {\it composite},
i.e. depend on $N$ in a very special way.
In this case, the HOMFLY polynomials need a more careful definition
(see {\it uniform polynomials} of \cite{MMkrM}),
and the example of Hopf link is a very good illustration.

\bigskip

Basically, there are three different types of expressions for the Hopf polynomials. The first one is due to M. Rosso and V. F. R. Jones \cite{RJ} and gives the HOMFLY invariant of the Hopf link in the form
\be
{\cal H}_{R\times S}^{\rm Hopf}=
q^{\varkappa_R+\varkappa_S} \sum_{Q\in R\otimes S}
N_{RS}^Q \cdot q^{-\varkappa_{_Q}}\cdot D_{_Q}(q )
\label{RJ}
\ee
where $N^{Q}_{RS}$ are the integer-valued Littlewood-Richardson coefficients
counting the multiplicity of representation ${Q}$ in the product of representations
$R$ and $S$, while
\be
\varkappa_Q = (\Lambda_Q,\Lambda_Q+2\rho)
\ee
is the corresponding eigenvalue of the second Casimir operator
and
\be\label{Weyl}
D_Q=\prod_{\alpha\in\Delta_+}{[\left(\Lambda_Q+\rho,\alpha\right)]\over [\left(\rho,\alpha\right)]}
\ee
is its quantum dimension.
Here $\Lambda_Q$ is the highest weight of the representation $Q$,
$\rho$ is the
Weyl vector, equal to the half sum of positive roots, and
square bracket denotes the quantum number:
\be
[x]=\frac{q^x-q^{-x}}{q-q^{-1}} = \frac{\{q^x\}}{\{q\}},\ \ \ \ \ \ \{x\}=x-x^{-1}
\ee
In the paper, we do not distinguish between the representation $R$ and the associated Young diagram: $R=[r_1,r_2,\ldots,r_{_{l_{\!_R}}}]$, where $l_{\!_R}$ is the number of columns in the Young diagram, and we always denote the diagrams with the capital letters, and their column heights with the small letters.

For $SU(N)$, the parameter $A=q^N$ captures all the dependence on $N$, except for the $U(1)$-factor $q^{2|R||S|\over N}$. This one arises because for arbitrary representation $R$  the second Casimir is equal to
\be\label{Cas}
\varkappa_R=2\kappa_R-{|R|^2\over N}+|R|N
\ee
with $\kappa_R=\sum_{r_{i,j}\in R}(j-i)$, where the sum goes over the boxes of the Young diagram $R$ and $\kappa_R$ is
the corresponding eigenvalue of the cut-and-join operator $\hat W_2$ \cite{MMN1},
\be
\hat W_2\ \Sch_Q = \kappa_Q\cdot\Sch_Q
\ee
where $\Sch_Q$ is the Schur function.

The second set of formulas for the Hopf polynomials are just very explicit finite sums considered in \cite{MMMtang}. This realization of the Hopf polynomials is most convenient in concrete calculations, but at the moment Hopf invariants are available in this form not in the most general case.

At last, there is the third kind of explicit formula for the Hopf polynomial, and it is the one which we discuss in this note:
\be
{\cal H}^{\rm Hopf}_{{S}\times{R}} =D_S\cdot \Sch_R\{p^{*{S}}\}
\label{Hopftimes00}
\ee
Here the Schur functions $\Sch_R\{p^{*{S}}\}$ are taken at special points $p^{*{S}}$ which can be considered as deformations of the {\it topological locus} \cite{MMM}
\be
p^*_k = \frac{[Nk]}{[k]} = \frac{\{A^k\}}{\{q^k\}}
\ee
appearing in the description of quantum (graded) dimensions
$D_{S} = \Sch_{S}\{p^*\}$.

The formulas of the type (\ref{Hopftimes00}) are known in other variables (in Miwa variables as opposed to the time variables $p_k$) since \cite{ML}, and they were later exploited in \cite{Mar}. The goal of this note to extend them to the case when both $R$ and $S$ are composite representations. This requires a proper extension of the topological locus $p^{*{S}}$ and a notion of the adjoint Schur functions.

Note that, throughout the paper, we only use the unreduced HOMFLY invariants.

\subsection*{Hopf recursion}

From

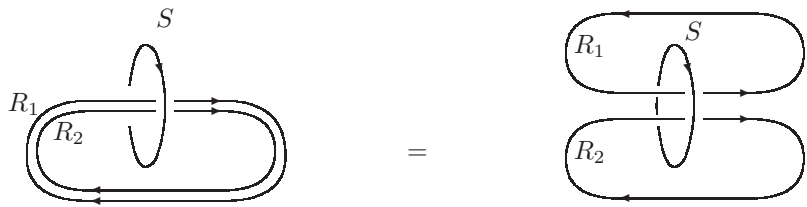
\begin{picture}(400,110)(-200,-60)

\put(-100,0){
\put(0,2){\line(1,0){25}}
\put(32,2){\vector(1,0){18}}
\put(25,-2){\line(-1,0){25}}
\put(32,-2){\vector(1,0){18}}

\qbezier(25,-20)(32,0)(25,20)
\qbezier(25,-20)(18,-30)(15,-8)
\qbezier(25,20)(18,30)(15,8)

\put(0,3){
\qbezier(0,-5)(-20,-5)(-20,-20)
\qbezier(0,-35)(-20,-35)(-20,-20)
\put(50,-35){\vector(-1,0){50}}
\qbezier(50,-5)(70,-5)(70,-20)
\qbezier(50,-35)(70,-35)(70,-20)
}

\qbezier(0,2)(-24,2)(-24,-19)
\qbezier(0,-36)(-24,-36)(-24,-19)
\put(50,-36){\vector(-1,0){50}}
\qbezier(50,2)(74,2)(74,-19)
\qbezier(50,-36)(74,-36)(74,-19)

\put(-31,-2){\mbox{${R}_1$}}
\put(-14,-13){\mbox{${R}_2$}}
\put(25,30){\mbox{${S}$}}
\put(25,20){\vector(1,-4){2}}

{\footnotesize
}
}

\put(100,0){
\put(0,5){\line(1,0){25}}
\put(32,5){\vector(1,0){18}}
\put(25,-5){\line(-1,0){25}}
\put(32,-5){\vector(1,0){18}}

\qbezier(25,-20)(32,0)(25,20)
\qbezier(25,-20)(18,-30)(15,-8)
\qbezier(25,20)(18,30)(15,8)
\qbezier(14.5,-3)(14.2,0)(14.5,3)

\qbezier(0,-5)(-20,-5)(-20,-20)
\qbezier(0,-35)(-20,-35)(-20,-20)
\put(50,-35){\vector(-1,0){50}}
\qbezier(50,-5)(70,-5)(70,-20)
\qbezier(50,-35)(70,-35)(70,-20)

\qbezier(0,5)(-20,5)(-20,20)
\qbezier(0,35)(-20,35)(-20,20)
\put(50,35){\vector(-1,0){50}}
\qbezier(50,5)(70,5)(70,20)
\qbezier(50,35)(70,35)(70,20)

\put(-17,20){\mbox{${R}_1$}}
\put(-17,-20){\mbox{${R}_2$}}
\put(25,25){\mbox{${S}$}}
\put(25,20){\vector(1,-4){2}}

\put(-80,-20){\mbox{$=$}}

{\footnotesize
}
}

\end{picture}

\noindent
it follows that
\be
\frac{{\cal H}^{\rm Hopf}_{{S}\times{R}_1}}{D_{S}}\cdot
\frac{{\cal H}^{\rm Hopf}_{{S}\times{R}_2}}{D_{S}}
=\sum_{{R}\in {R}_1\otimes{R}_2} N^{R}_{{R}_1{R}_2}\cdot
\frac{{\cal H}^{\rm Hopf}_{{S}\times{R}}}{D_{S}}
\ee
This equation resembles the group property of characters (Schur functions)
\be
\Sch_{{R}_1}\{p\}\cdot\Sch_{{R}_2}\{p\}
=\sum_{{R}\in {R}_1\otimes{R}_2} N^{R}_{{R}_1{R}_2}\cdot \Sch_{{R}}\{p\}
\ee
and implies that
\be
\frac{{\cal H}^{\rm Hopf}_{{S}\times{R}}}{D_{S}} \sim \Sch_R\{p^{*{S}}\}
\label{Hopftimes0}
\ee
with some ${S}$-dependent time-variables $p^{*{S}}$.
Relation (\ref{Hopftimes0}) fixes ${\cal H}^{\rm Hopf}_{{S}\times{R}}$ up to a factor with the multiplicative property $\xi (R_1)\xi (R_2)=\xi (R)$ for $R\in R_1\otimes R_2$. We fix it to be unity, which, with the proper choice of $p^{*S}$ (see below), fixes the framing of the Hopf invariant to be the standard, or canonical framing \cite{Atiah,MarF,China1,MMMtang} with the trivial topological $U(1)$ factor $U=q^{2|S||R|/N}$ taken into account\footnote{In practical terms, this framing is characterized by the quasiclassical expansion of the reduced link invariant, $q=e^\hbar$, $A=e^{N\hbar}$ without the linear term in $\hbar$:
\be
H=1+0\cdot\hbar+O(\hbar^2)
\nn
\ee
} (similarly with the Rosso-Jones formula (\ref{RJ})):
\be
{\cal H}^{\rm Hopf}_{{S}\times{R}} =D_S\cdot \Sch_R\{p^{*{S}}\}
\label{Hopftimes}
\ee

\bigskip

The symmetry
\be
{\cal H}^{\rm Hopf}_{{S}\times{R}} = {\cal H}^{\rm Hopf}_{{R}\times{S}}
\ee
implies a set of recursion restrictions in ${R}$ for the time variables in (\ref{Hopftimes}):
\be\label{locuseq}
\forall {S} \ \ \ \ \
\Sch_{S}\{p^{*{R}_1}\}\cdot \Sch_{S}\{p^{*{R}_2}\} =
\frac{D_{S}}{D_{{R}_1}D_{{R}_2}} \cdot
\sum_{R} N^{R}_{{R}_1{R}_2}\cdot D_{R} \cdot \Sch_{S}\{p^{*{R}}\}
\ee
For example, for ${S}=[1]$
\be\label{defp*}
p_1^{*{R}_1}p_1^{*{R}_2} =
\frac{D_{[1]}}{D_{{R}_1}D_{{R}_2}} \sum_{R} N^{R}_{{R}_1{R}_2} D_{R}\cdot p_1^{*{R}}
\ee

\bigskip

Equations (\ref{locuseq}) have the "trivial" $R$-independent solution
$p_k=p_k^*=\{A^k\}/\{q^k\}$, which respects the  equation's invariance
under the mirror map $(A,q)\longrightarrow (A^{-1},q^{-1})$.
It describes the quantum dimensions, which are of course a solution.
Remarkable, however, there are solutions, which spontaneously break mirror symmetry,
and thus appear in pairs.
One such solution describes the Hopf link and its mirror\footnote{
E.g. for ${R}_1=[m]$ and ${R}_2=[1]$, we have for (\ref{pstar0}):
\be
D_{[m]}p_1^{*[m]}p_1^{*{[1]}} = D_{[m+1]}p_1^{*[m+1]}+D_{[m,1]}p_1^{*[m,1]}
\ \ \ \ \ \Longrightarrow \ \ \ \ \ \ \ \ \ \ \ \ \ \ \ \ \ \ \ \ \ \ \ \ \ \ \ \ \ \
\nn \\
\Big(p_1^*+A\{q\}\Big)\left(p_1^*+\frac{(q^{2m}-1)A}{q}\right)D_{[m]}
= \left(p_1^*+\frac{(q^{2m+2}-1)A}{q}\right)\frac{\{Aq^m\}D_{[m]}}{\{q^{m+1}\}}
+ \left(p_1^*+\frac{(q^{2m}-1)A}{q}+\frac{(q^2-1)A}{q^3}\right)
\frac{[m]\{A/q\}D_{[m]}}{\{q^{m+1}\}}
\nn
\ee
}:
\be\label{pstar0}
\phantom{.}^{(1)}{p_k^{*R}} = p^*_k + A^k\sum_i q^{(1-2i)k}(q^{2kr_i}-1)
\ee
and
\be\label{pstar1}
\phantom{.}^{(2)}{p_k^{*R}} = \phantom{.}^{(1)}{p_k^{*R}}(A^{-1},q^{-1})
= p^*_k + A^{-k}\sum_i q^{(2i-1)k}(q^{-2kr_i}-1)
\ee
Hereafter, for the sake of definiteness, we use (\ref{pstar1}).

An important point here is that the solution (\ref{pstar1}) is again defined up to a factor $\xi(R)$: $p_k^{*R}\to \xi(R)^k\cdot p_k^{*R}$
with the multiplicative property $\xi (R_1)\xi (R_2)=\xi (R)$ for $R\in R_1\otimes R_2$, so that
\be
\Sch_S\{p^{*{R}}\}\to \xi(R)^{|S|}\cdot\Sch_S\{p^{*{R}}\}
\ee
if this factor is taken into account. We chose it to be
\be\label{U1}
\xi(R)=q^{2|R|/N}
\ee
i.e.
\be\label{pstar}
\boxed{
{p_k^{*R}} = q^{2|R|k\over N}\Big(p^*_k + A^{-k}\sum_i q^{(2i-1)k}(q^{-2kr_i}-1)\Big)
}
\ee
which is necessary for the correct accounting of the $U(1)$-factor in the framing of the Hopf invariant.
This will allow a smooth transition to the conjugate and composite representations below. Choosing instead a unit $\xi(R)=1$ would eliminate the $U(1)$-factor.


\bigskip

The power of the relation (\ref{Hopftimes}) is that, using
\be
{\cal H}^{\rm Hopf}_{{R}\times{S}} =D_R\cdot \Sch_S\{p^{*{R}}\}
\ee
one can deduce
the ${R}$-dependent, {\it but ${S}$-independent}
locus $p^{*{R}}=p^*+\delta p^{*{R}}$
from the study of {\it symmetric} representations ${S}=[s]$ only,
and then it remains the same for all ${S}$, not obligatory symmetric.


\subsection*{Generic expressions}

Explicit expressions for the Hopf polynomials with one of the components colored with the symmetric representation only are well known \cite{AENV,ArthAENV,MMMtang}:
\be
\left\{\begin{array}{rl}
{\cal H}^{\rm Hopf}_{[r]\times [s]} = & q^{2rs\over N}
D_{[r]}D_{[s]}\left(1 + \sum_{i=1}^{{\rm min}(r,s)} (-A)^{-i} q^{\frac{i(i+3)}{2}-i(r+s)}
\prod_{j=0}^{i-1} \frac{\{q^{r-j}\}\{q^{s-j}\}}{\{Aq^j\}}\right)
\\ & \\
{\cal H}^{\rm Hopf}_{[1^r]\times [s]} =& q^{2rs\over N}D_{[1^r]}D_{[s]}\left(1 - \frac{\{q^r\}\{q^s\}}{q^{s-r}A\{A\}}\right)
\\ &\\
{\cal H}^{\rm Hopf}_{R\times[s]} =&q^{2|R|s\over N}D_{R}D_{[s]}\left(
1 - q^{r_1+\ldots+r_l-s}[s](q^2-1)^2
\sum_{i=1}^l \frac{q^{-r_i+i-2}[l+r_i-i]!}{\prod_{j\neq i}^l
[r_i-r_j-i+j]}\cdot\sum_{k=0}^{r_i-i}\frac{q^{-2ks}(q^2-1)^k}{[r_i-i-k ]!\,
\prod_{j=r_i-i-k}^{r_i-i}A\{Aq^j\}}\right)
\end{array}\right.
\!\!\!\!\!\!\!\!\!\!\!\!\!\!\!\!\!\!\!\!\!\!\!\!\!\!\!
\ee
These formulas are reproduced from (\ref{Hopftimes}):
\be
\boxed{
{\cal H}^{\rm Hopf}_{R\times S}= D_R\cdot\Sch_S\{p^{*R}\} = D_S\cdot \Sch_R\{p^{*S}\}
\ \ \   {\rm with} \ \ \
p_k^{*R} = q^{2|R|k\over N}\Big(p_k^*- A^{-k} \sum_{i=1}^{l_R} q^{(2i-1-r_i)k}\{q^{kr_i}\}\Big)
}
\label{HopfRS}
\ee
However, when conjugate or composite representations are involved, one needs explicit {\it uniform}
formulas (see \cite{MMkrM}) for the locus $p^{*R}$ and the Schur functions $\Sch_S$ that do not depend on $N$ (for $sl_N$), though the representations themselves do.

When one considers ordinary representations $R$ and $S$, which are labeled by the same Young diagrams for all the $sl_N$ algebras at once, the issue of $U(1)$-factor is not very important and is often neglected. However, conjugate, adjoint and generic composite representations (see below) essentially depend on the choice of $N$, and, in this case, the $U(1)$-factor becomes crucially important.

\subsection*{Involving conjugate representation}

Despite the Young diagram associated with the conjugate representation explicitly depends on $N$ for $sl_N$, the Hopf polynomials involving the conjugate representations are very simple, they are based on the property \cite{MMMtang}
\be
{\cal H}^{\rm Hopf}_{\bar R\times S} = 
{\cal H}^{\rm Hopf}_{R,S}(A^{-1},q^{-1})
\label{barRS}
\ee
As an immediate corollary,
\be\label{1}
\!\!\!\!\!\!\!\!\!\!\!\!\!\!\!\!\!\!
\boxed{
\begin{array}{l}
{\cal H}^{\rm Hopf}_{\overline{R}\times S} \ \ \stackrel{(\ref{barRS})}{=}\
{\cal H}^{\rm Hopf}_{ {R}\times S}(A^{-1},q^{-1})
\ \stackrel{(\ref{HopfRS})}{=}\ D_{R}\cdot \Sch_S\{p^{*\overline{[R]}}\}
\ \ \ \ \ \ {\rm with} \\
\\
p^{*\overline{R}}_k =
p_k^{*R}(A^{-1},q^{-1}) = q^{-{2|R|k\over N}}\Big(p_k^*+A^k
\sum_{i=1}^{l_R} q^{(r_i+1-2i)k}\{q^{kr_i} \}\Big)
\end{array}
}
\ee
In fact, one can directly obtain this formula, (\ref{pstar0}) for $p^{*\overline{R}}_k$ from (\ref{pstar}) from the Young diagram describing the conjugate representation: $\bar R=[\underbrace{r_1,\ldots,r_1}_{N-l_{\!_R}},r_1-r_{_{l_{\!_R}}},r_1-r_{_{l_{\!_R}-1}},\ldots,r_1-r_2]$.

\subsection*{Involving adjoint}

The adjoint representation gives another example when the Young diagram, associated with the representation of $sl_N$, explicitly depends on $N$. In this case, also there is a uniform polynomial \cite{MMkrM}. It is equal to \cite{MMMtang}
\be\label{2}
\left\{\begin{array}{l}
{\cal H}_{ {\rm adj}\times[s]} = D_{[s]}\cdot \frac{\{A/q\}}{\{q\}^2}\cdot
\left(Aq^{2s} \{q\}  + \{A/q\} + \frac{\{q\}}{Aq^{2s}}
\right)
= D_{\rm adj} \cdot \Sch_{[s]}\{p^{*\rm adj}\}
 \\
 \\
{\cal H}^{\rm Hopf}_{{\rm adj}\times[s] }(A ,q^{-1})={\cal H}^{\rm Hopf}_{{\rm adj}\times [1^s] }(A,q)
= D_{\rm adj}  \cdot \Sch_{[1^s]}\{p^{*\rm adj}\}\\
\\
\boxed{
{\cal H}_{ {\rm adj}\times S} = D_{\rm adj}\cdot \Sch_S\{p^{*\rm adj}\}
\ \ \ \ \ \ \ \ {\rm with} \ \ \ \ \ \
p_k^{*\rm adj} = (q^{2k}-1+q^{-2k})\cdot p^*_k
}
\end{array}\right.
\ee
where again one reads  the explicit expression for $p_k^{*\rm adj}$ from (\ref{pstar}) for the adjoint representation $R=[2,\underbrace{1,\ldots,1}_{N-2}]$.

\subsection*{Involving composite representations}

Now we are ready to consider the general case of composite  (or rational, \cite{Koike,Kanno}; or coupled, \cite{Vafa}) representations $(R,P)$: the most general finite-dimensional irreducible highest weight representations of $sl_N$  \cite{Koike,GW,Vafa,Kanno,MarK}, which are associated with the Young diagram obtained by putting $R$ atop of
$p$ lines of the lengths $N-p_i^{\rm tr}$, i.e.
$$(R,P)= \Big[r_1+p_1,\ldots,r_{l_R}+p_1,\underbrace{p_1,\ldots,p_1}_{N-l_{\!_R}-l_{\!_P}},
p_1-p_{_{l_{\!_P}}},p_1-p_{{l_{\!_P}-1}},\ldots,p_1-p_2\Big]$$
or, pictorially,

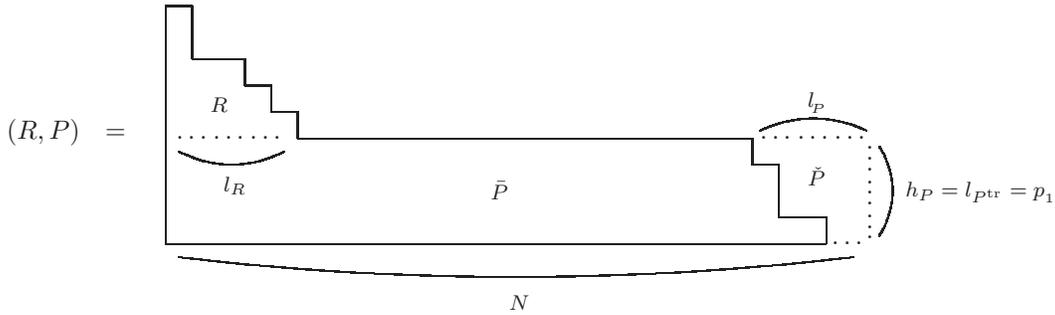
\begin{picture}(300,125)(-90,-30)

\put(0,0){\line(0,1){90}}
\put(0,0){\line(1,0){250}}
\put(50,40){\line(1,0){172}}

\put(0,90){\line(1,0){10}}
\put(10,90){\line(0,-1){20}}
\put(10,70){\line(1,0){20}}
\put(30,70){\line(0,-1){10}}
\put(30,60){\line(1,0){10}}
\put(40,60){\line(0,-1){10}}
\put(40,50){\line(1,0){10}}
\put(50,50){\line(0,-1){10}}

\put(265,2){\mbox{$\vdots$}}
\put(265,15){\mbox{$\vdots$}}
\put(265,28){\mbox{$\vdots$}}

\put(252,0){\mbox{$\ldots$}}
\put(253,40){\mbox{$\ldots$}}
\put(239,40){\mbox{$\ldots$}}
\put(225,40){\mbox{$\ldots$}}

\put(222,40){\line(0,-1){10}}
\put(222,30){\line(1,0){10}}
\put(232,30){\line(0,-1){20}}
\put(232,10){\line(1,0){18}}
\put(250,0){\line(0,1){10}}

\put(0,90){\line(1,0){10}}
\put(10,90){\line(0,-1){20}}
\put(10,70){\line(1,0){20}}
\put(30,70){\line(0,-1){10}}
\put(30,60){\line(1,0){10}}
\put(40,60){\line(0,-1){10}}
\put(40,50){\line(1,0){10}}
\put(50,50){\line(0,-1){10}}

\put(-60,40){\mbox{$(R,P) \ \ =$}}

{\footnotesize
\put(123,17){\mbox{$ \bar P$}}
\put(17,50){\mbox{$R$}}
\put(243,22){\mbox{$\check P$}}
\qbezier(270,3)(280,20)(270,37)
\put(280,18){\mbox{$h_P = l_{P^{\rm tr}}=p_{_1}$}}
\qbezier(5,-5)(132,-20)(260,-5)
\put(130,-25){\mbox{$N $}}
\qbezier(5,35)(25,25)(45,35)
\put(22,20){\mbox{$l_R$}}
\qbezier(225,43)(245,52)(265,43)
\put(243,52){\mbox{$l_{\!_P}$}}
}

\put(4,40){\mbox{$\ldots$}}
\put(18,40){\mbox{$\ldots$}}
\put(32,40){\mbox{$\ldots$}}

\end{picture}

\noindent
This $(R,P)$ is
the first ("maximal") representation, contributing to the product $R\otimes \bar P$. It can be manifestly obtained from the tensor products (i.e. as a projector from  $R\otimes \bar P$) by formula \cite{Koike}
\be
(R,P)=\sum_{Y,Y_1,Y_2}(-1)^{l_{\!_Y}}N^R_{YY_1}N^{P}_{Y^{\rm tr}Y_2}\ Y_1\otimes\overline{Y_2}
\ee
where the superscript "tr" denotes transposition. The quantum dimension of the composite representation is
\be\label{Dims}
\boxed{
D_{(R,P)}
=  D_{_R}(N-l_{\!_P})\, D_{_P}(N-l_{\!_R})\,
\frac{ \prod_{i=1}^{l_{\!_R}}[N-l_{\!P}-i]!\prod_{i'=1}^{l_{\!_P}}[N-l_{\!R}-i']!   }
{\prod_{i=1}^{l_{\!_R}+l_{\!_P} } [N-i]!} \, \prod_{i=1}^{l_{\!_R}}\prod_{i'=1}^{l_{\!_P}}
[N+r_i+p_{i'}+1-i-i']
}
\ee
where $D_R(N)$ is the quantum dimension of the representation $R$ of $sl_N$.

The adjoint representation in this notation is ${\rm adj} = ([1],[1])$,
while the conjugate of the representation $R$ is
$\overline{R} = (\emptyset,R)$.
The product
\be
[m]\otimes \overline{[m]} = \sum_{k=0}^m ([k],[k])
\ee
where $(\emptyset,\emptyset) \stackrel{sl_N}{\cong} \emptyset$,
and the other items are diagrams with $2k$ lines, $k$ of length $N-1$ and $k$ of length one.
Similarly,
\be
[1^m]\otimes \overline{[1^m]} = \sum _{k=0}^m ([1^k],[1^k])
\ee
where contributing are the diagrams with just two lines of lengths $N-k$ and $k$.

The general properties for the Hopf polynomials in the case of composite representations are
\be
{\cal H}^{\rm Hopf}_{(R,[1])\times S} =
{\cal H}^{\rm Hopf}_{ (R^{\rm tr},[1])\times S}(A,q^{-1})
\nn \\
{\cal H}^{\rm Hopf}_{(R,P)\times S} =
{\cal H}^{\rm Hopf}_{ (P,R)\times S}(A^{-1},q^{-1})
\label{RPPR}
\ee
A particular series is
\be
{\cal H}_{ ([r],[r])\times [s] }
= D_{([r],[r])}\cdot D_s\cdot  \left\{
1 + \{q\}^2\cdot \frac{[r][s] \{Aq^{r-1}\}\{Aq^{s}\}}{\{A\}\{Aq^{2r-1}\}}\cdot
\left(1\ + \ \sum_{i=1}^{r-1}\ \{q\}^i\cdot \frac{[r-1]!}{[r-1-i]!}\cdot
\frac{(A^{2i}q^{2is}+q^{-2is})}{\prod_{j=r-i}^{r-1} A\{Aq^j\}}
\right)
\right\} \nn
\ee
which is equivalent to
\be\label{3}
\boxed{
{\cal H}_{ ([r],[r])\times S} =
D_{([r],[r])}\cdot \Sch_S\{p^{_*([r],[r])}\}
\ \ \ \ \ \ \ \ {\rm with} \ \ \ \ \ \
p_k^{_*([r],[r])}
= p_k^{*{\rm adj}} + \{q^{k(r-1)}\}\{A^k q^{rk}\}
}
\ee
also following from (\ref{pstar}).

Explicit expression for the Hopf polynomial in arbitrary composite representation is far more involved, however, as usual it can be written in the form
\be\label{30}
{\cal H}_{ (R,P)\times S} =
D_{(R,P)}\cdot \Sch_S\{p^{_*(R,P)}\}
\ee
with  the most general Hopf topological locus $p^{_*(R,P)}$ implied by (\ref{pstar}) for the composite Young diagram $(R,P)$:
\be
\setlength\fboxrule{2pt}\setlength\fboxsep{2mm}
\boxed{
p_k^{*(R,P)} = q^{2{|R|-|P|\over N}k}\left(p_k^*+ \frac{1}{A^k}\cdot   \sum_{j=1}^{l_{\!_R}}  q^{(2j-1)k}\cdot(q^{-2kr_j}-1) + A^k\cdot  \sum_{i=1}^{l_P}q^{(1-2i)k}\cdot(q^{ 2kp_i}-1)
\right)
}
\label{compolocus}
\ee
This formula is in a perfect agreement with the previous examples (\ref{1}), (\ref{2}), (\ref{3}).

\subsection*{Adjoint Schur functions}

It remains to generalise expression (\ref{30}) to arbitrary composite representation $S$.
In order to calculate the Hopf polynomial in two composite representation, we need to simplify the expression for the Schur function for the composite representation. We call such Schur functions adjoint (they are associated with the universal characters of \cite{Koike} in the $gl_N$ case, see also \cite{Kanno}) and find here a simpler expression for them. We use the trick with the symmetricity of the Hopf polynomials w.r.t. interchanging colors of the components:
\be
 D_R\cdot \Sch_S\{p^{*{R}}\}={\cal H}^{\rm Hopf}_{{R}\times{S}}={\cal H}^{\rm Hopf}_{{S}\times{R}}=D_S\cdot \Sch_R\{p^{*{S}}\}
\ee
and calculate $\Sch_{(R,P)}\{p^{*{S}}\}$ via known $\Sch_S\{p^{*{(R,P)}}\}$.
The simplest example is
\be
{\cal H}^{\rm Hopf}_{{\rm adj}\times S} = D_{\rm adj}\cdot\Sch_S\{p^{*{\rm adj}}\}
= D_S\cdot\Sch_{\rm adj}\{p^{*S}\}   \ \ \ \ {\rm with} \ \ \ \
\boxed{\Sch_{\rm adj}\{p^{*S}\} = p_1^{*S}(A,q)\cdot p_1^{*S}(A^{-1},q^{-1}) - 1}
\nn
\ee
In general, the answer for
$\ {\cal H}^{\rm Hopf}_{(R,P)\times S} =
D_{(R,P)}\cdot\Sch_S\{p^{*(R,P)}\}
= D_S\cdot\Sch_{(R,P)}\{p^{*S}\}\ $ gives rise to
\be
\setlength\fboxrule{2pt}\setlength\fboxsep{2mm}
\boxed{
\Sch_{(R,P)}\{p^{*S}\}=
\sum_{\eta\in R\cap P^{\rm tr} } (-)^{|\eta|}\cdot\Sch_{R/\eta}\{p^{*S}\}
\cdot \Sch_{P/\eta^{\rm tr}}\{p^{*S}(A^{-1},q^{-1})\}
}
\label{compoSchur}
\ee
The sum  is automatically restricted to $\eta\in R\cap P^{\rm tr}$ by vanishing of the skew Schur polynomials $\Sch_{R/\eta}$ and $\Sch_{P/\eta^{\rm tr}}$ for bigger $\eta$. This formula immediately follows from (1) of Theorem 3.2 in \cite{Koike} and is eq.(4.19) of \cite{Kanno} given in time (not in Miwa) variables.

\subsection*{Hopf link with two composite representations}

The most general expression for the colored Hopf polynomial,
which complements the well-known (\ref{HopfRS}), is
\be
\setlength\fboxrule{2pt}\setlength\fboxsep{2mm}
\boxed{
 {\cal H}^{\rm Hopf}_{(R,P)\times (S,Q)} =
D_{(R,P)}\cdot\Sch_{(S,Q)}\{p^{*(R,P)}\}
= D_{(S,Q)}\cdot\Sch_{(R,P)}\{p^{*(S,Q)}\}
}
\label{compocompo}
\ee
where the composite Schur functions $\Sch_{(R,P)}\{p\}$
are defined in (\ref{compoSchur})
and the composite locus $p^{*(R,P)}$, in (\ref{compolocus}).

Expression (\ref{compocompo}) is the main issue of the present paper.
It has a natural extension to super- and hyper-polynomials,
which will be discussed elsewhere.

\subsection*{Appendix.}

In this Appendix, we  derive essential formulas (\ref{compoSchur}) and (\ref{Dims}).

\subsubsection*{On the derivation of (\ref{compoSchur})}

For the sake of completeness we present here a calculation illustrating obtaining formula (\ref{compoSchur}) for the general composite representation from (\ref{pstar}).
To simplify the formulas, we write them  only for $p_1$,
expressions for all other $p_k$
are obtained by a plethystic map provided by the substitution $(A,q)\longrightarrow (A^k,q^k)$.

As a preliminary exercise,
let us apply (\ref{pstar}) to the diagram $[h^N]$, which gives
\vspace{-0.2cm}
\be
q^{2h}\cdot\Big(p_1^ *+ A^{-1}\sum_{i=1}^N q^{2i-1}\cdot(q^{-2h}-1)\Big) = q^{2h}\cdot\Big(p_1^ *+
(q^{-2h}-1)\cdot A^{-1}q\cdot\Big(1+q^{2}+\ldots + q^{2N-2}\Big) =
\nn
\ee
\vspace{-1.0cm}
\be
\label{p1hN}
\ee
\vspace{-1.0cm}
\be
=q^{2h}\cdot\Big(p_1^* + (q^{-2h}-1)\cdot A^{-1}q \cdot\frac{1-q^{2N}}{1-q^{2}}\Big) =
q^{2h}\cdot\Big(\frac{A-A^{-1}}{q-q^{-1}} + (1-q^{1-2h})\cdot \frac{1-q^{2N}}{A(q-q^{-1})}\Big) =
\frac{A-A^{-1}}{q-q^{-1}} = p_1^*
\nn
\ee
where, at the last step, we substituted $q^{-2N}=A^{-2}$.
Since, for $sl_N$ algebra, all representations $[h^N]$ with arbitrary $h$ are
equivalent to the empty one, $[h^N] \stackrel{sl_N}{\cong}  \emptyset$,
the true time-variables should be just $p_k^*$, and we see that
they are actually reproduced by  (\ref{pstar}).

Let us now see how it works for a generic composite representation $(R,P)$,
and do the same calculation for the diagram

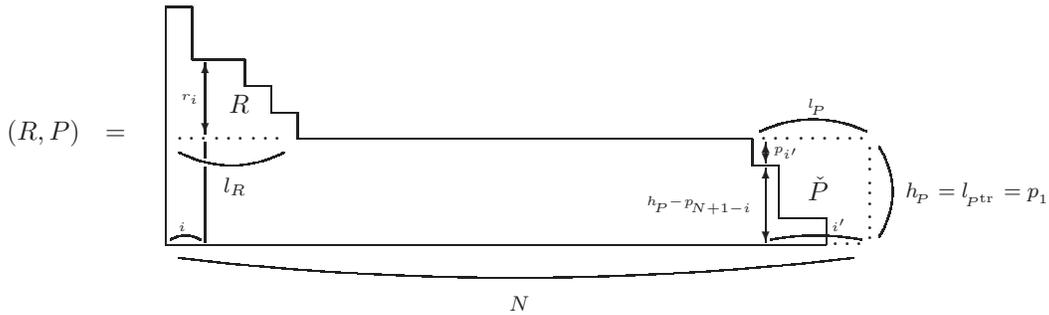
\begin{picture}(300,130)(-90,-30)

\put(0,0){\line(0,1){90}}
\put(0,0){\line(1,0){250}}
\put(50,40){\line(1,0){172}}

\put(0,90){\line(1,0){10}}
\put(10,90){\line(0,-1){20}}
\put(10,70){\line(1,0){20}}
\put(30,70){\line(0,-1){10}}
\put(30,60){\line(1,0){10}}
\put(40,60){\line(0,-1){10}}
\put(40,50){\line(1,0){10}}
\put(50,50){\line(0,-1){10}}

\put(265,2){\mbox{$\vdots$}}
\put(265,15){\mbox{$\vdots$}}
\put(265,28){\mbox{$\vdots$}}

\put(252,0){\mbox{$\ldots$}}
\put(253,40){\mbox{$\ldots$}}
\put(239,40){\mbox{$\ldots$}}
\put(225,40){\mbox{$\ldots$}}

\put(222,40){\line(0,-1){10}}
\put(222,30){\line(1,0){10}}
\put(232,30){\line(0,-1){20}}
\put(232,10){\line(1,0){18}}
\put(250,0){\line(0,1){10}}

\put(0,90){\line(1,0){10}}
\put(10,90){\line(0,-1){20}}
\put(10,70){\line(1,0){20}}
\put(30,70){\line(0,-1){10}}
\put(30,60){\line(1,0){10}}
\put(40,60){\line(0,-1){10}}
\put(40,50){\line(1,0){10}}
\put(50,50){\line(0,-1){10}}

\put(15,42){\vector(0,1){27}} \put(15,44){\vector(0,-1){2}}
\put(227,1){\vector(0,1){28}} \put(227,3){\vector(0,-1){2}}
\put(227,31){\vector(0,1){8}} \put(227,33){\vector(0,-1){2}}

\put(-60,40){\mbox{$(R,P) \ \ =$}}

\put(24,50){\mbox{$R$}}
\put(243,17){\mbox{$ \check P$}}

\qbezier(2,2)(7.5,5)(13,2) \put(15,1){\line(0,1){28.5}}\put(15,33){\line(0,1){6}}

{\footnotesize
\qbezier(270,3)(280,20)(270,37)
\put(280,18){\mbox{$h_{\!_P} = l_{_{\!P^{\rm tr}}}=p_{_1}$}}
\qbezier(5,-5)(132,-20)(260,-5)
\put(130,-25){\mbox{$N $}}
\qbezier(5,35)(25,25)(45,35)
\put(22,20){\mbox{$l_R$}}
}
{\tiny
\put(252,5){\mbox{$i'$}}  \qbezier(229,2)(246,5)(263,2)
\qbezier(225,43)(245,52)(265,43)
\put(243,52){\mbox{$l_{\!_P}$}}
\put(6,55){\mbox{$r_i$}}  \put(5,5){\mbox{$i$}}
\put(182,15){\mbox{$h_{\!_P}\!\!-\!p_{\!_{N+1-i}}$}}
\put(230,35){\mbox{$
p_{i'}$}}
}

\put(4,40){\mbox{$\ldots$}}
\put(18,40){\mbox{$\ldots$}}
\put(32,40){\mbox{$\ldots$}}

\end{picture}

\noindent
This time the sum in (\ref{pstar})
splits into three different pieces associated with
three different regions in the diagram (for simplicity, we assume that they do
not overlap, i.e. that $l_R+l_P\leq N$, but the answer remains the same if they do):
{\footnotesize
\be
\!\!\!\!\!\!\!\!
q^{2(|R|+Nh_P-|P|)\over N}\cdot\Big(p_1^* + A^{-1}\cdot \sum_{i=1}^{l_R}q^{2i-1}\cdot(q^{-2h_{\!_P}-2r_i}-1)
+A^{-1}\cdot\!\!\!\!\sum_{{i=l_R+1}}^{N-l_P}q^{2i-1}\cdot(q^{-2h_{\!_P}}-1)
+ A^{-1}\cdot\!\!\!\!\!\!\!\! \sum_{{i=N-l_P+1}}^N \!\!q^{2i-1}\cdot(q^{2p_{_{\!N+1-i}}-2h_{\!_P}}-1)\Big)
\nn
\ee}
(to avoid a confusion: $p$ in the last exponent is the height of the column in the diagram $P$,
not the time-variable).

\bigskip

We can now add $\ 0 = -q^{-2h_{\!_P}}+q^{-2h_{\!_P}}\ $ to each item and reshuffle the sum:
{\footnotesize
\be
q^{{2(|R|-|P|)\over N}+2h_P}\cdot\Big(
p_1^* + A^{-1}\cdot \sum_{i=1}^{l_R}q^{2i-1}\cdot(q^{-2h_{\!_P}-2r_i}-q^{-2h_{\!_P}}+q^{-2h_{\!_P}}-1)
+A^{-1}\cdot\!\!\!\!\sum_{i=l_R+1}^{N-l_P}q^{2i-1}\cdot(q^{-2h_{\!_P}}-1)+\nn\\
+ A^{-1}\cdot\!\!\!\!\!\!\!\! \sum_{i=N-l_P+1}^N \!\!\!\!
\!\!q^{2i-1}\cdot(q^{2p_{_{\!N+1-i}}-2h_{\!_P}}-q^{-2h_{\!_P}}+q^{-2h_{\!_P}}-1)\Big)
= \nn
\ee
\be
\!\!\!
= q^{{2(|R|-|P|)\over N}+2h_P}\cdot\Big(\underbrace{p_1^*
+ (q^{-2h_{\!_P}}-1)\sum_{i=1}^N q^{2i-1}}_{\stackrel{(\ref{p1hN})}{=}q^{-2h_{\!_P}}\cdot p_1^*}
+ \ q^{-2h_{\!_P}} \cdot A^{-1}\cdot\!\! \sum_{i=1}^{l_R}q^{2i-1}\cdot(q^{-2r_i}-1)+
+ \ q^{-2h_{\!_P}} \cdot A^{-1}\cdot\!\!\!\!\!\!\!\! \sum_{i=N-l_P+1}^N \!\!\!\!
\!\!q^{2i-1}\cdot(q^{ 2p_{_{\!N+1-i}}}-1)\Big)
\nn
\ee
}

\noindent
It remains to change the summation variable in the last sum $N+1-i=j$,
substitute $A\cdot q^{-2N} \stackrel{sl_N}{=} A^{-1}$ and restore $k$
 to get (\ref{compolocus})
\be
p_k^{*(R,P)} = q^{{2(|R|-|P|)\over N}}\cdot\Big(p_k^* + \frac{1}{A^k}\cdot   \sum_{i'=1}^{l_{\!_R}}  q^{(2i'-1)k}\cdot(q^{-2kr_{i'}}-1)+ A^k\cdot  \sum_{i=1}^{l_P}q^{(1-2i)k}\cdot(q^{ 2kp_i}-1)
\Big)
\ee

\subsubsection*{On the derivation of (\ref{Dims})}

In the same way, one can calculate the quantum dimensions of composite representations.
The quantum dimension of $sl_N$ representation associated with the Young diagram
$R=\{r_1\geq r_2\geq \ldots r_{l_{_R}}>0\}$
is given by the hook formula
\be
D_R = \prod_{(i,j)\in R} \frac{\{Aq^{j-i}\}}{\{q^{{\rm leg}_{(i,j)}+{\rm arm}_{(i,j)}+1}\}}
= \prod_{i=1}^{l_{_R}}\left( \frac{[N+r_i-i]!}{ [r_i-i+l_{\!_R}]!\, [N-i]!}
\prod_{j=i+1}^{l_{\!_R}} [r_i-r_j+j-i]\right)
\ee
For the composite representation, we get from this formula a combination of three products
coming from three "regions" of the diagram ${\cal R}=(R,P)$:
\be
{\cal R}_i = h_{\!_P} + r_i \ &{\rm for} & i=1,\ldots, l_{_R}
\nn\\
{\cal R}_i = h_{\!_P} \ \ \ \ \ \ \ \ \ \   & {\rm for} & i=l_{_R}+1,\ldots, N-l_{\!_P}
\nn\\
{\cal R}_i = h_{\!_P} - p_{N+1-i}\ & {\rm for} & N-l_{\!_P}+1,\ldots, N
\ee
Then

{\footnotesize
\be
\!\!\!\!\!\!\!\!\!\!\!\!\!
D_{(R,P)} =
\prod_{i=1}^{l_{\!_R}} \left(\frac{[N+h_{\!_P}+r_i-i]!}{ [h_{\!_P}+r_i-i+l_{\cal R}]!\, [N-i]!}
\prod_{j=i+1}^{l_{\!_R}} [h_{\!_P}+r_i-h_{\!_P}-r_j+j-i]
\prod_{j=l_{\!_R}+1}^{N-l_{\!_P} } [h_{\!_P}+r_i-h_{\!_P}+j-i]
\!\!\!\!\prod_{j=N-l_{\!_P} +1}^{l_{\cal R}=N} \!\!\!\!
[h_{\!_P}+r_i-h_{\!_P} + p_{N+1-j}+j-i]\right)\cdot
\nn
\ee
\vspace{-0.3cm}
\be
\cdot\prod_{i=l_{\!_R}+1}^{N-l_{\!_P}} \left(\frac{[N+h_{\!_P}-i]!}{ [h_{\!_P}-i+l_{\cal R}]!\, [N-i]!}
\prod_{j=i+1}^{N-l_{\!_P} } [h_{\!_P}-h_{\!_P}+j-i]
\!\!\!\!\prod_{j=N-l_{\!_P} +1}^{l_{\cal R}=N} \!\!\!\!
[h_{\!_P}-h_{\!_P} + p_{N+1-j}+j-i]\right)\cdot
\nn\\
\cdot \prod_{i=N-l_{\!_P}}^{N} \left(\frac{[N+h_{\!_P} - p_{N+1-i}-i]!}{ [h_{\!_P} - p_{N+1-i}-i+l_{\cal R}]!\, [N-i]!}
\prod_{j=i+1}^{l_{\cal R}=N}
[h_{\!_P} - p_{N+1-i}-h_{\!_P} + p_{N+1-j}+j-i]\right)
\nn
\ee
}

\noindent
Now we can cancel the differences of $h_{\!_P}$ and substitute $l_{\cal R}=N$,
which leads to further cancellations between the factorials in the numerators and denominators:

{\footnotesize
\be
D_{(R,P)} = \prod_{i=1}^N \frac{1}{[N-i]!}\cdot
\prod_{i=1}^{l_{\!_R}} \left(
\prod_{j=i+1}^{l_{\!_R}} [ r_i -r_j+j-i]
\prod_{j=l_{\!_R}+1}^{N-l_{\!_P} } [ r_i +j-i]
\!\!\!\!\prod_{j=N-l_{\!_P} +1}^{ N} \!\!\!\!
[ r_i  + p_{N+1-j}+j-i]\right)\cdot
\nn\\
\cdot\prod_{i=l_{\!_R}+1}^{N-l_{\!_P}} \left(
\prod_{j=i+1}^{N-l_{\!_P} } [ j-i]
\!\!\!\!\prod_{j=N-l_{\!_P} +1}^{ N} \!\!\!\!
[ p_{N+1-j}+j-i]\right)\cdot
\prod_{i=N-l_{\!_P}}^{N} \left(
\prod_{j=i+1}^{ N}
[ - p_{N+1-i}  + p_{N+1-j}+j-i]\right)   =
\nn \\
= \prod_{i=1}^N \frac{1}{[N-i]!}\cdot
\prod_{i=1}^{l_{\!_R}} \left(\frac{[N-l_{\!_P}+r_i-i]!}{[l_{\!_R}+r_i-i  ]!}
\prod_{j=i+1}^{l_{\!_R}} [ r_i -r_j+j-i]
 \prod_{j'=1}^{l_{\!_P}  }
[ r_i  + p_{j'}+N+1-j'-i]\right)\cdot
\nn\\
\cdot\prod_{i=l_{\!_R}+1}^{N-l_{\!_P}} \left(
  [ N-l_{\!_P}-i]!\cdot
\prod_{j'=1}^{l_{\!_P}}
[ p_{j'}+N+1-j'-i]\right)\cdot
\prod_{i'=1}^{l_{\!_P}}  \left(
\prod_{j'=i'+1}^{l_{\!_P}}
[ p_{i'}  - p_{j'}-i'+j']\right) =
\nn
\ee
\be
\!\!\!\!\!\!\!\!\!\!
= \frac{\prod_{i<j}^{l_{\!_R}} [r_i-r_j-i+j]
\cdot \prod_{i'<j'}^{l_{\!_P}} [p_{i'}-p_{j'}-i'+j'] }
{\prod_{i=1}^{|R|+|P| } [N-i]!}  \cdot
\prod_{i=1}^{l_{\!_R}}\frac{[N-l_{\!_P}+r_i-i]!}{[l_{\!_R}+r_i-i]!} \cdot
\prod_{i'=1}^{l_{\!_P}}\frac{[N-l_{\!_R}+p_{i'}-i']!}{[l_{\!_P}+p_{i'}-i']!}
\cdot
\prod_{i=1}^{l_{\!_R}}\prod_{i'=1}^{l_{\!_P}}
[N+r_i+p_{i'}+1-i-i']!
\nn
\ee
}

\bigskip

\noindent
Thus, finally we obtain (\ref{Dims}):

\be
D_{(R,P)}
=  D_{_R}(N-l_{\!_P})\, D_{_P}(N-l_{\!_R})\,
\frac{ \prod_{i=1}^{l_{\!_R}}[N-l_{\!P}-i]!\prod_{i'=1}^{l_{\!_P}}[N-l_{\!R}-i']!   }
{\prod_{i=1}^{l_{\!_R}+l_{\!_P} } [N-i]!} \, \prod_{i=1}^{l_{\!_R}}\prod_{i'=1}^{l_{\!_P}}
[N+r_i+p_{i'}+1-i-i']
\nn
\ee

\section*{Acknowledgements}

We appreciate illuminating discussions with H.Awata and H.Kanno.

Our work was supported by the Russian Science Foundation (Grant No.16-12-10344).


\end{document}